\documentclass[twocolumn,showpacs,showkeys,superscriptaddress]{revtex4}
\usepackage{graphicx}
\usepackage{amsmath}
\usepackage{amsfonts}
\usepackage{amssymb}
\usepackage{mathrsfs}

\def\openone{\leavevmode\hbox{\small1\kern-3.8pt\normalsize1}}
\def\N{\leavevmode\hbox{ Z \kern-8 pt\normalsize{Z}}}
\def\openone{\leavevmode\hbox{\small1\kern-3.8pt\normalsize1}}
\def\openJ{\leavevmode\hbox{J \kern-9.5pt\normalsize J}}
\def\openS{\leavevmode\hbox{ S \kern-9.3pt\normalsize S}}
\newcommand{\bb}{\begin{equation}}
\newcommand{\ee}{\end{equation}}
\newcommand{\eqb}{\begin{eqnarray}}
\newcommand{\eqf}{\end{eqnarray}}

\usepackage{color}

\begin{document}

\title{Spinning particles coupled to gravity}

\author{Sergio A. Hojman}
\email{sergio.hojman@uai.cl}
\affiliation{UAI Physics Center, Universidad Adolfo Ib\'a\~nez, Santiago, Chile.}
\affiliation{Departamento de Ciencias, Facultad de Artes Liberales,
Universidad Adolfo Ib\'a\~nez, Santiago, Chile.}
\affiliation{Departamento de F\'{\i}sica, Facultad de Ciencias, Universidad de Chile,
Santiago, Chile.}
\affiliation{Centro de Recursos Educativos Avanzados,
CREA, Santiago, Chile.}
\author{Felipe A. Asenjo}
\email{felipe.asenjo@uai.cl}
\affiliation{UAI Physics Center, Universidad Adolfo Ib\'a\~nez, Santiago, Chile.}
\affiliation{Facultad de Ingenier\'{\i}a y Ciencias,
Universidad Adolfo Ib\'a\~nez, Santiago, Chile.}

\begin{abstract}
Recent experimental work has determined that free falling $^{87}$Rb atoms on Earth, with vertically aligned spins, follow geodesics, thus apparently ruling out spin--gravitation interactions. It is showed that while some spinning matter models coupled to gravitation referenced to in that work seem to be ruled out by the experiment, those same experimental results confirm theoretical results derived from a Lagrangian description of spinning particles coupled to gravity constructed over forty years ago. A proposal to carry out (similar but) different experiments which will help to test the validity of the Universality of Free Fall as opposed to the correctness of the aforementioned Lagrangian theory, is presented.
\end{abstract}


\maketitle

\section{Introduction}

In a recently published Letter \cite{UFF} an experiment to asses the universality of free fall (UFF) by testing spin--gravity coupling was presented. The experiment considers free falling $^{87}$Rb atoms on Earth with vertically aligned spins pointing either up or down. The determination of the E\"otvos ratio for the accelerations of both kinds of spin orientation allows for the comparison of the experimental results with theoretical models of spin--curvature and spin--torsion couplings developed in \cite{spgr1,spgr2,spgr3}. The conclusion is that spin--curvature and spin--torsion couplings are not observed  at the level of $1.2 \times 10^{-7}$, thus disproving the aforementioned theoretical models. Nevertheless, we prove that those experimental results are exactly consistent with the ones predicted by a Lagrangian theory of spinning particles (tops) coupled to gravity constructed over forty years ago \cite{hojman1} and applied in different contexts over the years \cite{hojman3,hojman2a,hojman2,ha,hk,zha,abk,ahkz,nz,zgwyl}. The results obtained in \cite{hojman1} are developed starting from the Lagrangian flat spacetime formalism of spinning tops developed by Hanson and Regge in \cite{hr}.
Furthermore, we propose a similar (but different) experiment to test UFF against the spin--gravity coupling defined in this Lagrangian description of tops \cite{hojman1,hojman3,hojman2a,hojman2,ha,hk,zha,abk,ahkz,nz,zgwyl}. This test might yield a violation of UFF within the capabilities of the experimental setting as the one described in Ref.~\cite{UFF}.

\section{Lagrangian model for spinning particles}

 The dynamics of tops with mass $m$, spin $J$, energy $E$
and total angular momentum $j$ can be fully described in terms of a Lagrangian theory.
For a spinning particle, its velocity vector
$u^\mu$ and its canonical momentum vector $P^\mu$ are not parallel in general, and the
velocity vector may become spacelike \cite{hojman1,hr,hojman3} while the momentum vector remains timelike due to a dynamical conservation law of the (square of the) mass $m^2\equiv P^\mu P_\mu> 0$ \cite{hojman1,ha}. Usually, the spin of the particle is defined in terms of an antisymmetric tensor $ S^{\mu \nu}$.
The dynamics of a top is completely described by the non--geodesic equations of motion for the momentum \cite{hojman1,hojman2}
\begin{equation}
 \frac{D P^\mu}{D\lambda}=-\frac{1}{2}{R^\mu}_{\nu\alpha\beta}u^\nu S^{\alpha\beta}\, ,
\label{momentummotion}
\end{equation}
and for the spin tensor
\begin{equation}
\frac{D S^{\mu \nu}}{D\lambda}=S^{\mu
\lambda}{\sigma_\lambda}^\nu-\sigma^{\mu
\lambda}{S_\lambda}^\nu=P^\mu u^\nu-u^\mu P^\nu\, , \label{spinmotion}
\end{equation}
where $\sigma^{\mu \nu}$ is the antisymmetric angular velocity tensor. Here, $D_\lambda\equiv {D}/{D \lambda}$ is the covariant derivative, such that $ {D_\lambda P^\mu}\equiv \dot P^\mu+\Gamma^\mu_{\alpha\beta}P^\alpha u^\beta$, and
$ {D_\lambda S^{\mu\nu}}\equiv \dot S^{\mu\nu}+\Gamma^\mu_{\alpha\beta}S^{\alpha\nu} u^\beta+\Gamma^\nu_{\alpha\beta}S^{\mu\alpha} u^\beta$,
where the overdot represents the derivative with respect to an arbitrary parameter $\lambda$, and ${\Gamma^{\nu}}_{\rho \tau}$ are  the Christoffel symbols for the metric field $g_{\mu \nu}$ (the speed of light is set equal to $1$).
The six independent components of the antisymmetric spin tensor generate Lorentz transformations. In order to restrict them to describe three dimensional rotations only, the Tulczyjew constraint is usually imposed \cite{tulc,hojman1, hr}
\begin{equation}
S^{\mu \nu} P_\nu=0\, .\label{constraint}
\end{equation}

The behavior of a top moving on a background gravitational field is determined by Eqs.~\eqref{momentummotion},~\eqref{spinmotion}, and ~\eqref{constraint}. It is clear that the top does not follow geodesic paths [by the non-zero right-hand side of Eq.~\eqref{momentummotion}]. Thereby, the top can be understood as an extended object  that feels tidal forces due to gravity.
They are directly obtained from a Lagrangian formulation. The  position of the top is denoted by $x^\mu$, and its orientation is defined by an orthonormal tetrad ${e_{(\alpha)}}^{\mu}$ (with six independent components) \cite{hojman1,hojman2}. The orthonormality condition implies
$ g_{\mu \nu} \ {e_{(\alpha)}}^{\mu} \ {e_{{(\beta)}}}^{\nu} \ \equiv\ \eta_{(\alpha \beta)}$,
with $\eta_{(\alpha \beta)} (= \eta^{(\alpha \beta)})$ the flat-spacetime metric $\eta_{(\alpha \beta)} \ \equiv \ \ \mbox{diag}\ (+1, -1, -1, -1)$.
The top velocity vector $u^\mu$ is  thus defined in terms of the arbitrary
parameter $\lambda$ by
\begin{equation}
u^\mu\equiv \frac{d x^\mu}{d \lambda}\, ,\label{vel}
\end{equation}
whereas the antisymmetric angular velocity tensor is \cite{hojman1,hojman2}
\begin{equation}
\sigma^{\mu \nu}\ \equiv \eta^{(\alpha \beta)}
{e_{(\alpha)}}^{\mu}\frac{D{e_{{(\beta)}}}}{D \lambda} ^{\nu}\ = \ -
\ \sigma^{\nu \mu}\, ,\label{sigma}
\end{equation}
with${D{e_{{(\beta)}}}}^\nu/{D \lambda} \ \equiv \
{d{e_{{(\beta)}}}}^\nu/{d \lambda}\ + {\Gamma^{\nu}}_{\rho \tau} \
{e_{{(\beta)}}}^{\rho}\ u^\tau$.

Therefore, the action for the top dynamics $S=\int L\, d\lambda$, is chosen to be $\lambda$--reparametrization
invariant. The Lagrangian
\begin{equation}
L (a_1, a_2, a_3, a_4) = (a_1)^{1/2}\mathcal{L} \left(a_2/a_1,
a_3/(a_1)^2, a_4/(a_1)^2\right)\, , \label{lag1}
\end{equation}
is an
arbitrary function of four invariants $a_1, a_2, a_3, a_4$, and
$\mathcal{L}$ is an arbitrary function of $a_1\equiv u^\mu u_\mu,\ a_2\equiv \sigma^{\mu \nu} \sigma_{\mu \nu}=
- \mbox{tr}({\sigma}^2),\ a_3\equiv u_\alpha \sigma^{\alpha \beta}
\sigma_{\beta \gamma} u^\gamma,\ a_4\equiv \mbox{det}({\sigma})$, where
$u^\mu$ and $\sigma^{\mu \nu}$ are the top's velocity and angular velocity respectively.
The momentum vector $P_\mu$ and the antisymmetric spin tensor
$S_{\mu \nu}$ are canonically conjugated to the position and orientation of the top. They are defined by
\begin{equation}
P_\mu \equiv \frac{\partial L}{\partial u^\mu}\, ,\qquad
S_{\mu \nu} \equiv \frac{\partial L}{\partial \sigma^{\mu \nu}} = -
S_{\nu \mu}\, .\label{sigmamunu}
\end{equation}
Without a  Lagrangian formulation for a top,  the canonical momentum cannot be
appropriately defined.
The non--geodesic equations of motion \eqref{momentummotion} and \eqref{spinmotion}
can be obtained by variation of
the action $S$ (for an arbitrary  $\cal L$) with respect to ten independent  $\delta
x^\mu$ and the covariant generalization of $\delta \theta^{\mu \nu} \equiv \eta^{(\alpha \beta)}
{e_{(\alpha)}}^{\mu}{\delta {e_{{(\beta)}}}}^\nu\ = - \delta
\theta^{\nu \mu}$.

The consistency of the constraint \eqref{constraint} with the
equations of motion \eqref{momentummotion} and \eqref{spinmotion} is
guaranteed making use of the arbitrariness of Lagrangian \cite{hr}. This implies that the Tulczyjew constraint can be considered as a dynamical property of the arbitrary Lagrangian, and not an external imposition on the top dynamics. In Ref.~\cite{hr,ha} an explicit Lagrangian function has been built to give rise to equations of motion \eqref{momentummotion} and \eqref{spinmotion} and to the constraint \eqref{constraint}.

Furthermore, it is possible to show that both the
top mass $m$ and its spin $J$ are conserved quantities in this theory \cite{ha}
\begin{equation}
m^2\equiv P^\mu P_\mu\, , \qquad
J^2\equiv \frac{1}{2} S^{\mu \nu}S_{\mu \nu}\, . \label{spin}
\end{equation}
Lastly, if $\xi^\mu$ is a Killing vector, then
\begin{equation}\label{killingC}
C_\xi\equiv P^\mu \xi_\mu-\frac{1}{2}S^{\mu \nu}\xi_{\mu;\nu}\, ,
\end{equation}
is a constant of motion \cite{hojman1,hojman2,hojman3}.

\section{Free fall with vertically aligned spin}

The experiment described in Ref.~\cite{UFF} consists in a free falliing top with its spin aligned (parallel or antiparallel) to its vertical trajectory.
In this section we show that the theory presented above agrees exactly with the results of Ref.~\cite{UFF}.

Let us consider the Earth's Schwarzschild field. In order to better model a free fall, let us write the metric in cartesian coordinates $x^0=t,x,y,z$ such that $g_{00}=1-2 r_0/r$, where $2 r_0$ is the Schwarzschild radius, and $r=\sqrt{x^2+y^2+z^2}$. Similarly $g_{0i}=0$, and $g_{ij}=-\delta_{ij}-2 r_0 x^i x^j/(r^3-2r_0 r^2)$ \cite{hojman1}.

As we model a free fall in this  gravitational field, then we set $x=\dot x=0$ and $y=\dot y=0$ as initial conditions. It is a straightforward matter to prove that these conditions are preserved b y the dynamics so that the particle only moves along the $z$-direction. Therefore, the momentum is aligned along the free-fall direction $P^x=0=P^y$. In Ref.~\cite{UFF}, the spin is chosen to be along the direction of the top motion, then $S^{xy}\neq 0$ is the only non-zero spin component. We show that dynamics defined by the previous assumptions are consistently allowed by  Eqs.~\eqref{momentummotion}--\eqref{constraint}.

First, it is straightforward to prove that the four constraints \eqref{constraint} are identically satisfied by our choices. On the other hand, the spin equation \eqref{spinmotion} for the ${0z}$-components yields the relation
\begin{equation}
P^0 u^z= P^z u^0\, ,
\end{equation}
implying that the momentum along the free-fall direction is proportional to the velocity in that direction (similar to the spinless case). Also, the spin equation for ${x y}$-components turns out to be
\begin{equation}\label{freefallspinconserv}
\dot S^{xy}=0\, ,
\end{equation}
and then, the spin in the $z$-direction is conserved along the trajectory. All other components for the spin equations are identically satisfied. By Eq.~\eqref{spin}, we can identify $S^{xy}=\pm J$ as the two possible orientations of the spin of the particle \cite{explain}.

Finally, the time component of Eq.~\eqref{momentummotion} yields
\begin{equation}\label{momfree0}
\frac{D P^0}{D\lambda}\equiv \dot P^0+2\Gamma^0_{0z}P^0 u^z=0\, ,
\end{equation}
whereas the $z$-component becomes
\begin{equation}\label{momfreez}
\frac{D P^z}{D\lambda}\equiv \dot P^z+\Gamma^z_{00}P^0 u^0+\Gamma^z_{zz}P^z u^z=0\, .
\end{equation}
In both Eqs.~\eqref{momfree0} and \eqref{momfreez} the spin gravity coupling vanish identically [the right-hand side in Eq.~\eqref{momentummotion} vanishes]. Similarly, the $x$ and $y$-components vanish identically.

The above solution describes a free falling top where the spin is initially orientated along the direction of motion. Eq.~\eqref{freefallspinconserv} establish that the spin vector of the particle will remain constant along the whole motion, and that any measurement of the momentum or velocity of the particle will only reflect the dynamics of a geodesic motion in free fall [Eqs.~\eqref{momfree0}-\eqref{momfreez}]. Thereby,  the experiment performed in Ref.~\cite{UFF}  is agrees exactly with this Lagrangian theory on a Schwarzschild background.

The particular solution detailed above is the one where the spin decouples from the gravitational field. To obtain a solution where the spin-gravity coupling be relevant, a different trajectory should be studied.

\section{``Parabolic'' motion with spin perpendicular to the plane of motion}

Let us assume Schwarzschild field background to dewscribe the Earth gravitational field. In this case, the equatorial motion of a top can be solved exactly (notice that due to spherical symmetry there are infinite many equatorial planes defined by each of the vertical planes where the ``parabolic'' motion takes place). We go back to write the metric in spherical coordinates for simplicity $x^i=t,r,\theta,\phi$.
Thus, the metric is $g_{00}=1-2r_0/r$,
$g_{rr}=-\left(1-2r_0/r\right)^{-1}$, $g_{\theta\theta}=-r^2$,
$g_{\phi\phi}=-r^2\sin^2\theta$.

Without any loss of generality, we can study the the motion in the
plane defined by $\cos\theta=0$. If the top is initially in that
plane and $\dot\theta=0$, then it remains in the equatorial plane
\cite{hojman1}, in which $\theta=\pi/2$ and $P^\theta=0$. Also the spin can be chosen to be orthogonal to the equatorial plane $S^{r\theta}=S^{\theta\phi}=S^{0\theta}=0$. Thus, the spin remain parallel to the angular momentum of the top along the trajectory.
We refer the reader to Ref.~\cite{ha} where this solution is fully developed.
Here, we limit ourselves to exhibit the most important features of this solution. Solving Eqs.~\eqref{momentummotion}, using the Killing vector conservation laws \eqref{killingC}
we get \cite{ha}
\begin{equation}
 P_\phi=\frac{-j\pm E J/m}{1-\eta}\, ,\qquad
 P_t=\frac{E\mp j J r_0/(m r^3)}{1-\eta}\, ,
\label{pt}\end{equation}
and
\begin{equation}
 P^r= \left[{P_t^2}-\left(\frac{P_\phi^2}{r^2}+m^2\right)\left(1-\frac{2r_0}{r}\right)\right]^{1/2}\, ,
\label{pr1}\end{equation}
which is obtained  from $P_\mu P^\mu=m^2$. Here,  $E$ is the top's energy and  $j$ is its total angular
momentum. Also we have defined the dimensionless parameter
\begin{equation}
 \eta=\frac{J^2 r_0}{m^2 r^3}\, ,
\label{eta}
\end{equation}
where $J$ is the top's spin given by Eq.~\eqref{spin}.
Now, using Eq.~\eqref{spinmotion} in the plane
$\theta=\pi/2$,
\begin{eqnarray}\label{}
  \frac{D S^{tr}}{D\lambda}=P^t \dot r-P^r\, ,\qquad
  \frac{D S^{t\phi}}{D\lambda}=P^t \dot \phi-P^\phi\, ,
\end{eqnarray}
althogeter with the relations \cite{ha}
\begin{equation}
S^{tr}=-\frac{S^{\phi r}P_\phi}{P_t}\, ,\quad
S^{t\phi}=\frac{S^{\phi r}P_r}{P_t}\, ,\quad
\left(S^{\phi r}\right)^2=\frac{J^2 \left(P_t\right)^2}{m^2 r^2}\, ,
\end{equation}
derived from Eqs.~\eqref{constraint} and \eqref{spin}, will let us find the solution for the velocities  \cite{ha}
\begin{equation}
 \dot\phi=\frac{1}{r^2}\left(1-\frac{2 r_0}{r}\right)\left(\frac{2\eta+1}{\eta-1}\right)\left(\frac{P_\phi}{P_t}\right)\, ,
\label{phipunto}\end{equation}
\begin{equation}
 \dot r=\left(1-\frac{2 r_0}{r}\right)\left(\frac{P^r}{P_t}\right)\, .
\label{rpuntop1}\end{equation}
\begin{widetext}
The problem is completely solved. However, we are interested in any correction to the trajectories that tops follow. Using above solution we can get \cite{ha,hojman1}
\begin{equation}
\frac{d\phi_\pm}{dr}= \frac{\left(2\eta+1\right)}{\left(1-\eta\right)^2}\left(\frac{j\mp EJ/m}{r^2}\right)\left[\frac{(E\mp j J r_0/mr^3)^2}{(1-\eta)^2}-\left(1-\frac{2r_0}{r}\right)\left(m^2+\frac{(-j\pm EJ/m)^2}{(1-\eta)^2 r^2}\right) \right]^{-1/2}\, .
\label{dphidr}
\end{equation}
\end{widetext}

The above trajectory yields the usual results for geodesic motion in the Schwarzschild field when $J=0$ ($\eta=0$) \cite{hartle}. Also, if the particle is freely fallingl with $P_\phi=0$ and $\dot\phi=0$, then the top has a nonzero total angular momentum $j=\pm EJ/m$. Thus, $P_t=E$ and $P^r$ becomes again the classical radial momentum for the geodesic motion in the Schwarzschild field.

It is clear that the equatorial plane trajectory \eqref{dphidr}  contains as some of its solutions the``parabolic'' motion of the top. However at this point is clear that for any `` parabolic'' trajectory there are two different trajectories described by the term $j\mp E J/m$ in \eqref{dphidr}. These two trajectories depend on the spin orientation, parallel or antiparallel to the total angular momentum of the top, remaining both of them perpendicular to the plane of motion.

The spin coupling to gravity also affects the acceleration of the top. From Eqs.~\eqref{pt} and \eqref{rpuntop1} we can readily obtain
\begin{eqnarray}\label{accelerationPtphi}
\dot P_t &=&\mp \frac{3J r_0}{m r^4 (1-\eta)}\left(1-\frac{2r_0}{r}\right)\frac{P_\phi P^r}{P_t}\, ,\nonumber\\
\dot P_\phi &=&\mp \frac{J}{m}\dot P_t\, .
\end{eqnarray}
Similarly, from Eq.~\eqref{pr1} we can get the radial top component for the acceleration
\begin{eqnarray}\label{accelerationPrfreefall}
\dot P^r&=&\frac{P_t \dot P_t}{P^r}-\frac{P_\phi \dot P_\phi}{P^r r^2}\left(1-\frac{2r_0}{r}\right)\nonumber\\
&&+\frac{1}{P_t}\left(1-\frac{2r_0}{r}\right)\left[\frac{P_\phi^2}{r^3}\left(1-\frac{3 r_0}{r}\right)-\frac{r_0 m}{r^2} \right]\, .
\end{eqnarray}

Spinless particles experience a radial force only. For a top, however, its spin, and its interplay with gravity, introduces corrections to the radial acceleration $\dot P^r/m$, and a new acceleration $\dot P^\phi/m$ in the $\phi$-direction. One possible way to detect these effects is presented in the following section.

\section{Estimations}

 ``Parabolic'' motion is one of the best candidates to be find spin--gravity coupling. Free falling tops with spin vertically aligned behave as spinless particles. But in a ``parabolic'' motion the spin orientation plays a crucial role in the non-geodesic motion of the top.

Let us assume a top in an experiment near the Earth surface, such that $r\sim R\gg r_0$ (where $R$ is the Earth radius), and $J\sim \hbar$. Also assume $j\gg J$,  $\eta \ll 1$, and neglect ${\cal O}(\hbar^2)$. In this case we have $P_\phi\approx -j$, $P_t\approx E\approx m$, where we have considered that the initial velocity of the top much smaller than the speed of light.

First, form Eq.~\eqref{dphidr} we can see that the top's trajectory differences for the two types of spin orientations
can be estimated as
\begin{equation}\label{eotvostraj}
\eta_\phi=\frac{d_r\phi_--d_r\phi_+}{d_r\phi_-+d_r\phi_+}\propto\frac{J}{j}\sim 6\times 10^{-20}\left(\frac{m_e}{m}\right)\left(\frac{c}{v}\right)\, ,
\end{equation}
where we have approximated the total angular momentum $j\sim m v R$, where $v$ is the initial top velocity, and $m_e$ is the electron mass (we have reinserted $c$ for the sake of clarity). If initially the top velocity is of the order of [mm/s], then $\chi_\phi\sim 10^{-8} (m_e/m)$. Therefore, for an electron, this trajectory difference could, in principle, be detected with current experimental capabilities.

On the other hand, one can also wonder about the differences on the top's acceleration between the two spin orientations in a ``parabolic'' motion. Using our approximations, Eqs.~\eqref{accelerationPtphi} reduce to
\begin{equation}
\dot P_t\approx \pm \frac{3 J j r_0}{m^2 r^4}{P^r}\, ,\qquad \dot P_\phi\approx 0\, ,
\end{equation}
and the top has a radial acceleration given by
\begin{equation}
a_{\pm=}\frac{\dot P^r}{m}\approx -\frac{r_0}{r^2}+\frac{j^2}{m^2 r^3}\pm\frac{3 J j r_0}{m^2 r^4}
\end{equation}
where the first term corresponds to the acceleration of gravity.  We can calculate the E\"otv\"os ratio for the top motion acceleration
\begin{eqnarray}\label{eotvos}
\eta_a=\frac{a_--a_+}{a_-+a_+}\approx \frac{3 J j}{m^2 c^2 r^2}\left(1-\frac{j^2}{m^2 c^2 r_0 r}\right)^{-1}\, .
\end{eqnarray}
If we assume that $v\ll c\sqrt{r_0/R}$ (the Earth's escape velocity), then  near the Earth surface ($r=R\gg r_0$) we have
\begin{eqnarray}\label{eotvos2}
\eta_a\approx \left(\frac{3 J}{m_e c R}\right)\left(\frac{m_e}{m}\right)\left(\frac{v}{c}\right)\sim 10^{-19}\left(\frac{m_e}{m}\right)\left(\frac{v}{c}\right)\, .
\end{eqnarray}
For this case, the  E\"otv\"os ratio is very small ranging outside the current experimental capabilities.
 However, the ratio \eqref{eotvos} can increase if the top total angular momentum approaches the critical value $j_0= m c \sqrt{r_0 R}$, which corresponds to a top velocity of the order of the Earth's escape velocity.

The estimations \eqref{eotvostraj} and \eqref{eotvos2}  predict that a ``parabolic'' motion for a top must show some deviations from a classical geodesic motion, whereas its free-falling motion \eqref{momfree0} and \eqref{momfreez} will not present any difference from the spinless dynamics. It is in the parabolic motion where the spin-gravity coupling can be observed. We hope that this results encourage the search for these effects in general relativity.

\begin{acknowledgments}
F.A.A. thanks the CONICyT-Chile for partial support through Funding No. 79130002.
\end{acknowledgments}

\end{document}